\documentclass[12pt]{article}
\usepackage{latexsym}
\usepackage{graphicx}
\begin{document}
\title{Focusing of world-lines in Weyl gravity}
\author{Morteza Mohseni\thanks{Corresponding author. Email: m-mohseni@pnu.ac.ir}, Mohsen Fathi
\\{Physics Department, Payame Noor University, Tehran 19395-3697, Iran}}

\maketitle

\begin{abstract}
We study the evolution of time-like congruences in the vacuum solutions of Weyl conformal theory of gravity. Using the Raycaudhuri equation, we show that for 
positive values of the coefficient of the linear term in the solution and in the absence of the cosmological constant, the incoming rays converge. The evolution of 
the congruence for negative values is investigated for different values of the parameters. The behavior of the congruence under conformal transformations is also studied.
\\
PACS: 04.20Cv, 04.50Kd
 \end{abstract}
\section{Introduction}\label{intro}
The existence of space-time singularities is a common property of most solutions of the general theory of relativity, including various black-hole or cosmological ones.
The well-known Raychaudhuri equation \cite{Raychaudhuri3} has played a key role in describing such singularities \cite{Penrose,Hawking1}. This equation has also been used 
in different other contexts \cite{alb,rip,kar2,shoj,har,grec,val,borg,stav,tsag,alb2,tsagas,kou,moh}. It has also attracted some attention beyond the classical descriptions, for example, 
a Bohmian quantum version of the equation was recently introduced in Ref.\cite{bohm}. 

In the present work, we aim to investigate this equation in the framework of the Weyl conformal theory of gravity by considering time-like world-lines. This theory emerged from 
attempts to unify gravity with electromagnetism and since its advent in about a century ago, it continued to attract the attention of researchers in the field. Sharing several important
solutions with the Einstein theory of gravity, it also possesses the interesting property of being invariant under conformal transformations, see Ref. \cite{Mannheim} and references 
therein. The relationships between this theory and the general theory of relativity have been discussed in several places, namely, in Ref. \cite{mald}, in which Maldacena has 
shown that it is possible to single out the Einstein solution from the numerous solution to Weyl theory by deploying a simple Neumann boundary condition, and in Ref. \cite{gote},
where it has been shown that varying the full connection of the Weyl gravity results in the vacuum Einstein equations. Null trajectories in Weyl gravity have been studied 
in \cite{villa}. The connections between Weyl gravity and extensions of the critical gravity was considered in \cite{lu}. Some difficulties with Weyl gravity have been discussed in 
Refs. \cite{flag,yoon}. 

Here, we consider the vacuum solution of the Weyl gravity which is a three-parameter Schwarzschild-like solution supplemented by linear and quadratic terms. This solution has been used
in different proposals, say, in \cite{Mannheim} to explain the galactic rotation curves, in \cite{sultan} to study the Sagnac effect, in \cite{azim} to investigate strong lensing, and in 
\cite{magnet} to study gravito-magnetic effects. The classical issues of bending of light and perihelion precession have been re-examined with this theory in Refs. \cite{kazana,sofie}
and \cite{democ}, respectively.

Our motivations for the present study originates from both the interest in Raychaudhuri equation and its applications in different contexts, and the Weyl conformal gravity as a theory 
of gravity with conformal invariance. The conformal invariance in gravity theories is regarded as an essential symmetry, see e.g. the discussion in \cite{thooft}. 
Also, modified gravities with quadratic curvature have attracted a lot of attention in recent decade and Weyl gravity as a particular model in this class, deserves further study 
in this regard.
 
In what follows, we start with the action and field equations of the Weyl conformal gravity. We then proceed by a brief review of the vacuum static solution of the theory and 
relevant equations for time-like geodesics. Using these equations, we apply the Raychaudhuri equation to a congruence of radial flow and a flow with both radial and rotational 
velocities to investigate the focusing of geodesics. After comparing the results with those of the standard Schwarzschild space-time, we investigate the effect of conformal 
transformations. A summary of the results will conclude the work.

\section{Time-like Geodesics in Weyl Gravity}
The Weyl theory of gravity is a theory of fourth order with respect to the metric. It is characterized by the action
\begin{equation}\label{9}
S=-\kappa\int{d^4x}\sqrt{-g}\,\,C^2,
\end{equation}
where $$C^2=C_{\mu\nu\rho\lambda}C^{\mu\nu\rho\lambda}$$ is the Weyl invariant, and $\kappa$ is a coupling constant. 
Using the properties of the Gauss-Bonnet invariant, the above action can be rewritten as
\begin{equation}\label{11}
S=-\kappa\int{{d}^4x}\sqrt{-g}\,\,\left(R^{\mu\nu}R_{\mu\nu}-\frac{1}{3}R^2\right).
\end{equation}
Varying the action given in Eq. (\ref{11}), supplemented by a matter action, with respect to $g_{\mu\nu}$ one gets the following field 
equation
\begin{equation}\label{13}
W_{\mu\nu}=\frac{1}{4\kappa} T_{\mu\nu},
\end{equation}
in which 
\begin{eqnarray}\label{12}
W_{\mu\nu}&=&\nabla^\rho\nabla_\mu R_{\nu\rho}+\nabla^\rho\nabla_\nu
R_{\mu\rho}-\Box R_{\mu\nu}-g_{\mu\nu}\nabla_\rho\nabla_\lambda
R^{\rho\lambda}\nonumber\\&&-\frac{1}{3}\Big(2\nabla_\mu\nabla_\nu
R-2g_{\mu\nu}\Box R-2RR_{\mu\nu}+\frac{1}{2}g_{\mu\nu}R^2\Big)\nonumber\\&&-2R_{\rho\nu}
R^{\rho}_\mu+\frac{1}{2}g_{\mu\nu}R_{\rho\lambda}R^{\rho\lambda}
\end{eqnarray}
is the Bach tensor \cite{Mannheim}, and $T_{\mu\nu}$ is the energy-momentum tensor. The vacuum field equation 
\begin{equation}\label{a1}
W_{\mu\nu}=0
\end{equation}
admits a static spherically symmetric solution defined by the line element 
\begin{equation}\label{14}
ds^2=-B(r)dt^2+B^{-1}(r)dr^2+r^2d\Omega^2,
\end{equation}
in which
\begin{equation}\label{15}
B(r)=-\frac{\beta(2-3\beta\gamma)}{r}+(1-3\beta\gamma)+\gamma r-kr^2.
\end{equation}
This solution was first introduced in Ref. \cite{Mannheim}. By choosing appropriate values for the parameters $\beta, \gamma$, and
$k$, the Schwarzschild-de Sitter metric could be regenerated. The parameters $\beta$ and $k$ can be related to the
mass of the source and the cosmological constant, respectively. In fact, $\beta(2-3\beta\gamma)$ gives the source mass.  
The third parameter, $\gamma$ is thus the crucial one here, carrying additional physical content,
which might be related to dark matter \cite{Mannheim}.

\begin{figure}[h]
\centering
\includegraphics{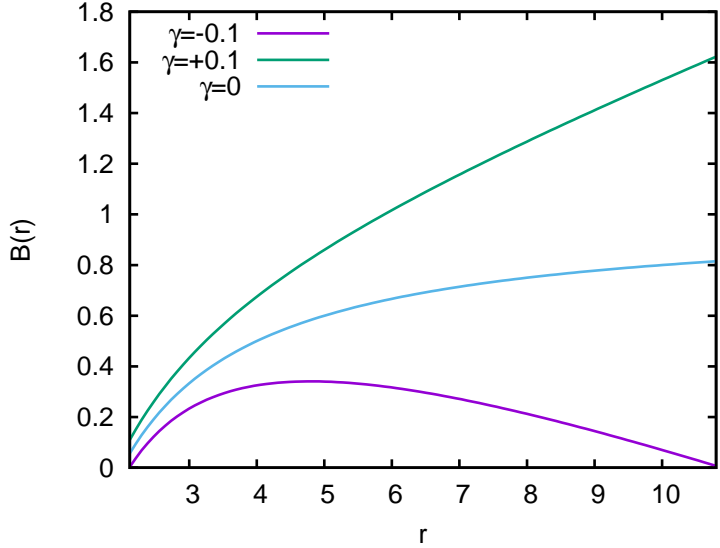}
\caption{Plot of $B(r)$ in terms of $r$ for $\beta=1$; $\gamma=-0.1$ (lower), 
$\gamma=0$ (middle), and $\gamma=+0.1$ (upper).}
\label{Fig0}
\end{figure}

For the the space-time described by the metric (\ref{14}) and (\ref{15}), if we take $k=0$, then depending on the sign of $\gamma$, there 
are one or two horizons corresponding to the roots of $B(r)$. These are given by 
$$r_{1,2}=\frac{3\beta}{2}+\frac{-1\pm\sqrt{1+\beta\gamma-2\beta^2\gamma^2}}{2\gamma}.$$ By assuming $\beta\gamma\ll 1$, we obtain
\begin{eqnarray}
r_1&\approx &2\beta\left(1-\frac{1}{2}\beta\gamma\right),\label{r1}\\
r_2&\approx &\beta(1+\beta\gamma)-\frac{1}{\gamma}\label{r2}	
\end{eqnarray}
which are consistent if $\gamma<0$. If, in addition, $|\gamma|\ll 1$, then $r_2\approx -\frac{1}{\gamma}$. For $\gamma>0$, the second root 
is ruled out, and one is left with the horizon corresponding to $r_1\approx 2\beta$. The function $B(r)$ is plotted in Fig. \ref{Fig0}
for both typical positive and negative values of $\gamma$. As this figure shows, in the case where $\gamma<0$, there is a maximum whose 
location is given by $r_0=\frac{\sqrt{-\beta\gamma(2-3\beta\gamma)}}{-\gamma}$, and the maximum value is given by
\begin{equation}
B_0\equiv B(r_0)=1-3\beta\gamma-2\sqrt{-\beta\gamma(2-3\beta\gamma)}
\end{equation}    
Taking $\beta=1$, these have the numerical values $r_0\approx 4.80, 14.25, 44.75$; and $B_0\approx 0.34, 0.75, 0.91$ for 
$\gamma=-0.1, -0.01, -0.001$, respectively.

Now, we study the evolution of time-like geodesics in the space-time defined by Eq. (\ref{14}). 
We start with the 4-velocity vector field
\begin{equation}\label{17}
v^\mu=\left(\dot t(\tau), \dot r(\tau), \dot\theta(\tau), \dot\phi(\tau)\right),
\end{equation}
with $v_\mu v^\mu =-1$, which defines a time-like congruence in the space-time under consideration. Here, $\dot x^\mu$ stands for
$\frac{dx^\mu}{d\tau}$, $\tau$ being an affine parameter along the relevant integral curves.
Due to the symmetries of the space-time described by Eq. (\ref{15}), we can set $\theta=\frac{\pi}{2}$. Then, the time-like Killing 
vectors $K=\partial_t$ and $L=\partial_\phi$ together with the fact that $K_\mu\frac{dx^\mu}{d\tau}=const.$ (see, e.g., \cite{car}), 
result in 
\begin{eqnarray}
{\dot t}&=&\frac{a}{B(r)},\label{econ1}\\
{\dot\phi}&=&\frac{b}{r^2}\label{econ2}
\end{eqnarray}
where $a,b$ are constants. Thus, from the geodesic equation
\begin{equation}
{\ddot r}-\frac{1}{2}{\dot r}^2(\ln B(r))^\prime+\frac{1}{2}{\dot t}^2B(r)B^\prime(r)={\dot\phi}^2rB(r),\label{eq2}
\end{equation}
or, from the normalization of the four-velocity, one obtains
\begin{equation}
\dot r=\pm\sqrt{a^2-B(r)-\frac{b^2}{r^2}}.\label{23}
\end{equation}
For a radial flow, $b=0$, we have
\begin{equation}
\dot r=\pm\sqrt{\frac{a^2r-3 \beta ^2 \gamma +2 \beta +k r^3+3 \beta  \gamma  r-\gamma  r^2-r}{r}},\label{24}
\end{equation}
In Schwarzschild space-time, for $r\rightarrow\infty$, we obtain ${\dot t}=a$. Thus, by setting $a=1$, one can 
choose ${\dot t}=1$, as viewed by an observer at infinity. In fact, it is well-known that $a$ is equal to the particle energy in units 
of $mc^2$.

In the space-time under consideration here, which is not asymptotically flat, we can not simply start by choosing $a=1$. To fix the 
constant $a$, one needs to use some initial conditions, say, the initial value of $\dot r$, or the maximum distance the particles can reach. 

The choice of the value of $a$ affects the motion of particles. This can be seen from Eq. (\ref{24}). For, $b=0$ and $\gamma<0$, if we 
choose $a^2\leq B_0$, there will be turning points whose locations, $r_m$, are given by the solutions of the equation ${\dot r}=0$, or 
equivalently, $a^2-B(r)=0$, and the particles motion are confined to $r_1<r\leq r_m<r_0$, or $r_0<r_m\leq r<r_2$, depending on the initial 
positions. If we choose $a>\sqrt{B_0}$, there will be no turning points and particles can pass through $r_0$. It would be useful to note 
that $\sqrt{B_0}$ is equal to, for example, about $0.58, 0.86, 0.96$ for $\gamma=-0.1, -0.01, -0.001$, respectively.  
         
\section{Raychaudhuri equation}
We now use the well-known Raychaudhuri equation 
\begin{equation}\label{1}
\frac{d\Theta}{d\tau}+\frac{1}{3}
\Theta^2+\sigma^2-\omega^2=-R_{\mu\nu} v^\mu v^\nu,
\end{equation}
to study the behavior of the above congruence. Here, $\Theta=\nabla_\mu v^\mu$, $\sigma_{\mu\nu}=
\nabla_{(\nu}v_{\mu)}-\frac{1}{3}h_{\mu\nu}\Theta$, and $\omega_{\mu\nu}=\nabla_{[\nu}v_{\mu]}$ are the scalar expansion, the symmetric 
traceless shear tensor, the antisymmetric rotation tensor, respectively,  $h_{\mu\nu}=g_{\mu\nu}+v_\mu v_\nu$, $\sigma^2=\sigma_{\mu\nu
}\sigma^{\mu\nu}$, and $\omega^2=\omega_{\mu\nu}\omega^{\mu\nu}$.

\begin{figure}[h]
\centering
\includegraphics{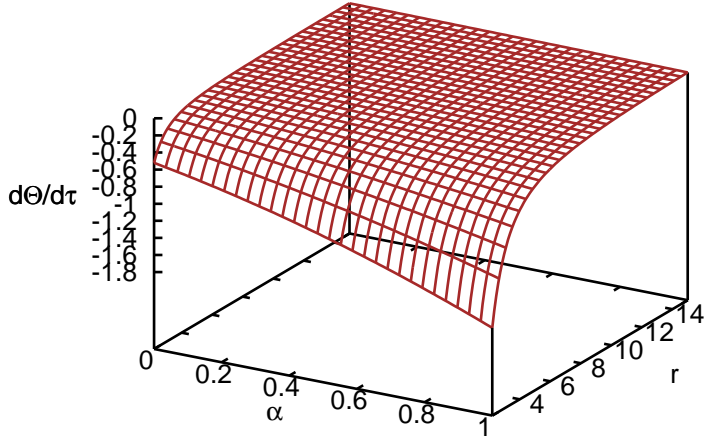}
\caption{Plot of $\frac{d\Theta}{d\tau}$ in terms of $r$ and $\alpha$ for $\beta=1, \gamma=-0.01$.}
\label{Fig4}
\end{figure}

For the radial congruence described by Eq. (\ref{24}), it can be shown that the rotation tensor vanishes and therefore,
\begin{equation}\label{27}
\frac{d\Theta}{d\tau}=-\frac{1}{3}\Theta^2-\sigma^2-\frac{\gamma}{r}
\end{equation}
The right hand side of this equation is negative for $\gamma\geq 0$, which means focusing of the congruence. However, for $\gamma<0$, 
the right hand side is not necessarily negative and the equation does not guarantee convergence. To analyze the latter case, we use a direct 
calculation of the expansion. By neglecting the cosmological constant, we obtain
\begin{eqnarray}\label{26}
\Theta=\mp\frac{5\gamma  r^2-3 \beta  (2-3 \beta  \gamma)+4r(1-a^2-3\beta\gamma)}{2 r^{3/2} \sqrt{-\gamma  r^2+\beta  (2-3 \beta  \gamma
		)+(a^2-1+3\beta\gamma)r}}\nonumber\\
\end{eqnarray}
in which the negative sign corresponds to outgoing world-lines ${\dot r}=\sqrt{a^2-B(r)}$, and the positive sign is associated with incoming
ones with ${\dot r}=-\sqrt{a^2-B(r)}$. Now, the rate of change of this expression can be obtained either by direct differentiation, or by
using the right hand side of Eq. (\ref{27}). The resulting expression have a rather algebraically complicated numerator quartic in $r$, and 
factors of $a^2-B(r)$ in the denominator. Thus, for an arbitrary value of $a$, there are up to four sign changes (corresponding to the 
roots of the numerator), and a divergency (in either side of $r_0$, corresponding to the roots of the expression in the denominator). For clarity, we consider
the case where there are no turning points, which is the case for $a>\sqrt{B_0}$. For convenience, we take $a^2=B_0+\alpha$ which means 
$\alpha>0$. The function $\frac{d\Theta}{d\tau}$ is plotted in Fig. \ref{Fig4} in terms of $r$ and $\alpha$ for an incoming congruence with a 
typical value $\gamma=-0.01$. This shows that the world-line are convergent. 

To have a better understanding of the physical content of Eq. (\ref{26}), we consider the galaxy cluster Abell 370 which is known 
particularly in relation with strong lensing phenomenon \cite{abell}. It has a mass of $M(<250 kpc)=3.8\times 10^{14}M_{\odot}\sim 7.6
\times 10^{44} kg$. Using this, we may set $\beta=\frac{GM}{c^2}\sim 5.6\times 10^{17} m$. Thus, we have $\frac{\beta(2-3\beta\gamma)}{r}
\sim 1.5\times 10^{-4}$, $\gamma r\sim 7.7 \times 10^{-5}$, i.e., the linear term and $\frac{\beta}{r}$ term in $B(r)$ are of the same 
order. Also, $\gamma r^2\sim 5.9\times 10^{17} m$, while other terms (except $\beta$) in the numerator of $\Theta$ in Eq. (\ref{26}) are 
some orders of magnitude smaller, that is, the main contribution to $\Theta$ comes from the first two terms in the numerator. 

Pure rotational flow is also possible with zero expansion as for Schwarzschild space-time. For combined radial-rotational flow 
(in the equatorial plane), we have from Eq. (\ref{23})
\begin{equation}
\dot r=\pm\sqrt{\frac{a^2r^3+(b^2+r^2)F(r)}{r^3}}\label{24b}
\end{equation} 
in which $$F(r)=-3 \beta ^2 \gamma +2 \beta +k r^3+3 \beta  \gamma  r-\gamma  r^2-r.$$ 
One can show that the relevant 
rotation tensor vanishes for this flow, and the Raychaudhuri equation reduces to
\begin{equation}\label{27a}
\frac{d\Theta}{d\tau}=-\frac{1}{3}\Theta^2-\sigma^2-\frac{\gamma}{r}\left(1+\frac{b^2}{r^2}\left(1+\frac{3\beta}{r}\right)\right)
\end{equation}
again, leading to convergent congruence for non-negative $\gamma$. For negative $\gamma$, the above equation does not result in convergence necessarily. In fact, in this case we have
\begin{equation}\label{26a}
\Theta=\mp\frac{F_1(r)}{2 r^{5/2}\sqrt{F_2(r)}}
\end{equation}
where
\begin{eqnarray*}
F_1(r)&=&5\gamma r^4+4(1-a^2-3\beta\gamma)r^3\\&&+3(b^2\gamma-\beta(2-3\beta\gamma))r^2\\
&&+2b^2(1-3\beta\gamma)r+b^2\beta(2-3\beta\gamma)\\	
F_2(r)&=&-\gamma  r^4+(a^2-1+3\beta\gamma)r^3\\&&+\beta(2-3\beta\gamma)r^2-b^2(1-3\beta\gamma)r\\&&+b^2\beta(2-3\beta\gamma).
\end{eqnarray*}
This results in a rather complicated expression for $\frac{d\Theta}{d\tau}$, and general behavior of the congruence depends on both $a,b$.

For radial null geodesics, the geodesic equation results in
\begin{equation}\label{n1}
v^\mu=\left(\frac{a}{B(r)},\pm a,0,0\right).
\end{equation}
By choosing $a=1$, the Raychaudhuri equation reads
\begin{equation}\label{27n}
\frac{d\Theta}{d\tau}=-\frac{1}{2}\Theta^2
\end{equation} 
that is, focusing with no signature of $\gamma$. This has the solution $\Theta=-\frac{2}{r}$.
\section{Conformal Transformations}
The action for Weyl gravity and the associated vacuum field equation are invariant under the conformal transformation
\begin{equation}\label{13a}
{\bar g}_{\mu\nu}=e^{2\varphi}g_{\mu\nu}
\end{equation}
where $\varphi$ is a scalar field. Thus, any metric related to the vacuum solution of the field equations via conformal transformations is 
also a solution. However, time-like geodesics are not necessarily mapped to geodesics. Thus, one can ask to what extent results obtained from 
equations like Raychaudhuri equation in a given space-time are valid in the transformed space-time. In this section, we aim to investigate 
the behavior of time-like congruences and the Raychaudhuri equation under such transformation. 

We start with
\begin{equation}\label{13b}
{\bar v}^\mu=e^{-\varphi}v^\mu
\end{equation} 
and 
\begin{equation}\label{13c}
{\bar\Gamma}^\mu_{\alpha\beta}=\Gamma^\mu_{\alpha\beta}+\delta^\mu_\alpha\partial_\beta\varphi+\delta^\mu_\beta\partial_\alpha\varphi
-g_{\alpha\beta}\partial^\mu\varphi
\end{equation} 
from which it can be shown that
\begin{equation}\label{13d}
\frac{\bar D{\bar v}^\mu}{{\bar D}\tau}=e^{-2\varphi}\frac{Dv^\mu}{D\tau}+e^{-2\varphi}\frac{d\varphi}{d\tau}v^\mu
+\frac{1}{2}(v_\nu v^\nu)\partial^\mu(e^{-2\varphi}).
\end{equation}
For null geodesics, the last term in the right-hand side of the above equation disappears and one is left with the geodesic equation
written in non-affine form. This is a re-statement of the well-known fact that null geodesics remain geodesics under a conformal 
transformation. For time-like geodesics, this is not the case. 

Regarding the Raychaudhuri equation for time-like congruences, we have
\begin{eqnarray}\label{13e}
{\bar\Theta}&=&{\bar\nabla}_\alpha{\bar v}^\alpha\nonumber\\
&=&e^{-\varphi}(\Theta+3{\dot\varphi})
\end{eqnarray}    
or
\begin{equation}\label{13f}
\frac{d\bar\Theta}{d\tau}=e^{-\varphi}({\dot\Theta}-\Theta{\dot\varphi}+3{\ddot\varphi}-3{\dot\varphi}^2).
\end{equation}  
This means that, in general, there is no guarantee that a convergent congruence remains convergent in the transformed space-time.  

\begin{figure}[h]
\centering
\includegraphics{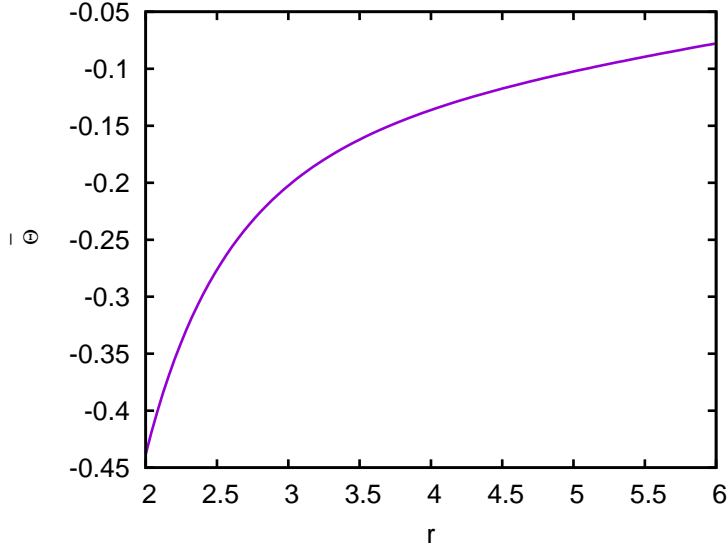}
\caption{Plot of $\bar\Theta$ in terms of $r$ for , $\varphi(r)=B(r)$, $\beta=1$, $a=1,\gamma=-0.1$}
\label{Fig2}
\end{figure}

For a scalar field $\varphi(r)$ which depends only on the radial coordinate, Eq. (\ref{13f}) reduces to 
\begin{equation}\label{13g}
\frac{d\bar\Theta}{d\tau}=e^{-\varphi}({\dot\Theta}-\Theta{\dot r}\varphi^\prime-3{\dot r}^2{\varphi^\prime}^2+3\varphi^\prime{\ddot r}
+3{\dot r}^2{\varphi}^{\prime\prime})
\end{equation}   
where a prime stands for differentiation with respect to the radial coordinate. As an example, if we choose $\varphi(r)=B(r)$, we obtain and 
$v^\mu=\left(\frac{a}{B(r)},-{\sqrt{a^2-B(r)},0,0}\right)$, which leads to function $\bar\Theta$ plotted in Fig. \ref{Fig2}. It shows that
the congruence is convergent in this particular transformed space.

The above results show how the trajectories behave in the transformed space. It is also possible to have a form of Raychaudhuri equation
unchanged when transforming to a conformally related space. This can be reached by considering a dynamical particle mass \cite{mancon}.  
\section{Conclusions}
In conclusion, we studied the focusing of time-like congruence within the space-time
described by the static vacuum solution to the Weyl conformal gravity model. For galactic scales, in which we ignored the cosmological 
constant, a term proportional to $\frac{-\gamma}{r}$ is added to the right-hand side of the Raychaudhuri equation,
where the rotation term is already absent. For positive $\gamma$ this ensures the convergence of incoming geodesics with a rate greater 
than the case of the Schwarzschild space-time. We also obtained a relation between the rate of convergence of a congruence for a given 
solution and in its conformally transformed space-time. The resultant equations do not guarantee that a convergent congruence remains so 
in the transformed space. In the context of cosmology, these results might help
to study how cosmic flows will behave in Weyl conformal gravity.

\section*{Acknowledgments}
We would like to thank Payame Noor University for financial support.

\end{document}